\documentclass[preprint,showpacs,preprintnumbers,amsmath,amssymb]{revtex4}

\usepackage{graphicx}
\usepackage{dcolumn}
\usepackage{bm}

\begin{document}


\title{$^8$He nuclei stopped in nuclear track emulsion}

 \author{D.~A.~Artemenkov}
   \affiliation{Joint Insitute for Nuclear Research, Dubna, Russia} 
 \author{A.~A.~Bezbakh}
   \affiliation{Joint Insitute for Nuclear Research, Dubna, Russia}
 \author{V.~Bradnova}
   \affiliation{Joint Insitute for Nuclear Research, Dubna, Russia} 
 \author{M.~S.~Golovkov}
   \affiliation{Joint Insitute for Nuclear Research, Dubna, Russia}
 \author{A.~V.~Gorshkov}
   \affiliation{Joint Insitute for Nuclear Research, Dubna, Russia}    
 \author{G.~Kaminsky}
   \affiliation{Institute of Nuclear Physics, Polish Academy of Sciences, Krakow, Poland}     
  \author{S.~A.~Krupko}
   \affiliation{Joint Insitute for Nuclear Research, Dubna, Russia}  
 \author{N.~K.~Kornegrutsa}
   \affiliation{Joint Insitute for Nuclear Research, Dubna, Russia}
  \author{R.~R.~Kattabekov}
   \affiliation{Institute for Physics and Technology, Uzbek Academy of Sciences, Tashkent, Republic of Uzbekistan} 
 \author{K.~Z.~Mamatkulov}
   \affiliation{Djizak State Pedagogical Institute, Djizak, Republic of Uzbekistan} 
 \author{V.~V.~ Rusakova}
   \affiliation{Joint Insitute for Nuclear Research, Dubna, Russia} 
  \author{R.~S.~Slepnev}
   \affiliation{Joint Insitute for Nuclear Research, Dubna, Russia}     
  \author{R.~Stanoeva}
   \affiliation{SouthWest University, Blagoevgrad, Bulgaria}  
 \author{S.~V.~Stepantsov}
   \affiliation{Joint Insitute for Nuclear Research, Dubna, Russia}  
 \author{A.~S.~Fomichev}
   \affiliation{Joint Insitute for Nuclear Research, Dubna, Russia}  
 \author{V.~Chudoba}
   \affiliation{Institute of Physics, Silesian University in Opava, Czech Republic }    
 \author{P.~I.~Zarubin}
     \email{zarubin@lhe.jinr.ru}    
     \homepage{http://becquerel.jinr.ru}
   \affiliation{Joint Insitute for Nuclear Research, Dubna, Russia} 
 \author{I.~G.~Zarubina}
   \affiliation{Joint Insitute for Nuclear Research, Dubna, Russia}   

\date{\today}

\begin{abstract}
\indent  The  fragment separator ACCULINNA in the G. N. Flerov Laboratory of Nuclear Reactions of JINR  was used to expose
 a nuclear track emulsion to a beam of radioactive $^{8}$He nuclei of energy of 60 MeV and enrichment of about 80\%.
 Measurements of decays of $^{8}$He  nuclei stopped in the emulsion allow one to evaluate possibilities of $\alpha$-spectrometry and
 to observe a thermal drift of $^{8}$He  atoms in matter. Knowledge of the energy and emission angles of $\alpha$-particles allows one to derive the energy distribution of $\alpha$-decays Q$_{2\alpha}$. The presence
 of a "tail" of large values Q$_{2\alpha}$ is established. The physical reason for the appearance of this "tail"
 in the distribution  Q$_{2\alpha}$ is not clear. Its shape could allow one to verify calculations of spatial structure of nucleon ensembles
 emerging as $\alpha$-pairs of  decays via the state $^8$Be$_{2+}$.\par
\end{abstract}

\maketitle
\section{}
\indent At the energy of a few MeV per nucleon, there is a possibility to study decays of radioactive nuclei  by implanting them into a detector   \cite{Hyldegaard_1,Hyldegaard_2,Hyldegaard_3,Mianowski_OTPC}.
 In particular, population of 2$\alpha$- and 3$\alpha$-particle states is possible in decays of light radioactive nuclei. In this respect,
 the unique, although somewhat forgotten, possibilities of nuclear track emulsion (NTE) for the detection of slow radioactive nuclei are worthy
 to be mentioned. The advantages of this method are the best spatial resolution (about 0.5 $\mu$m), the possibility of observing the tracks in a full
 solid angle and a record sensitivity starting with relativistic singly charged particles with a minimum ionization. In NTE, the directions and ranges
 of the beam nuclei, as well as slow products of their decays can be measured, which provides a basis for 
spectrometry. More than half a
 century ago, hammer-like tracks from the decay of $^8$Be nuclei through the first excited state 2$^+$ of about 2.0 MeV were observed in NTE.
 They occurred in the $\beta$-decays of stopped $^8$Li and $^8$Be fragments, which in turn were produced by high-energy particles \cite{Powell59}.
 Another example is the first observation of the $^9$C nucleus from the decay 2$\alpha$ + p \cite{Swami}. When used with sufficiently pure secondary beams,
 NTE appears to be an effective means for a systematic study of the decay of light nuclei with an excess of both neutrons and protons.\par
\begin{figure}
\centering
  \includegraphics[width=5in]{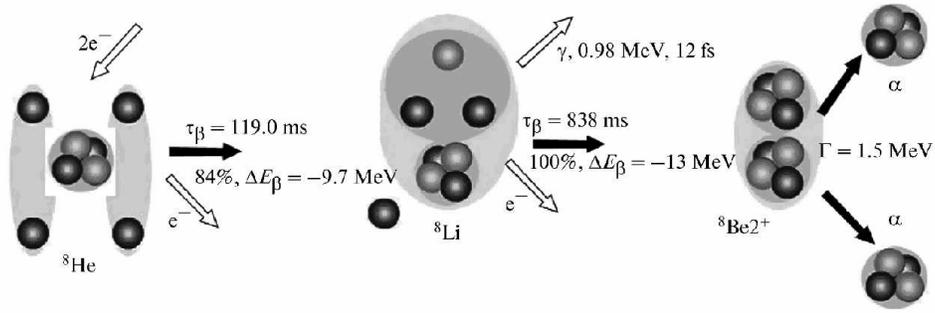}
\caption{ \label{fig:1}   Scheme of a major channel of the cascade decay of $^8$He isotope; light circles correspond to protons, dark ones -neutrons.}
  
\end{figure}

 \indent  In March 2012 exposure of NTE to nuclei $^8$He of energy of 60 MeV \cite{Artemenkov_8He} is performed at the fragment separator ACCULINNA \cite{Rodin} in the G. N. Flerov Laboratory
 of Nuclear Reactions, JINR. Features of decays of the $^8$He isotope are shown in Fig.~\ref{fig:1}, according to the compilation \cite{Ajzenberg}.
 Fig. 2 shows a mosaic macrophotograph of a decay of a nucleus $^8$He stopped in NTE. It is typical one among
 thousands observed in this study. Video recordings
 of such decays taken with the microscope and camera are collected \cite{Web2013}.\par
\begin{figure}
\centering
  \includegraphics[width=5in]{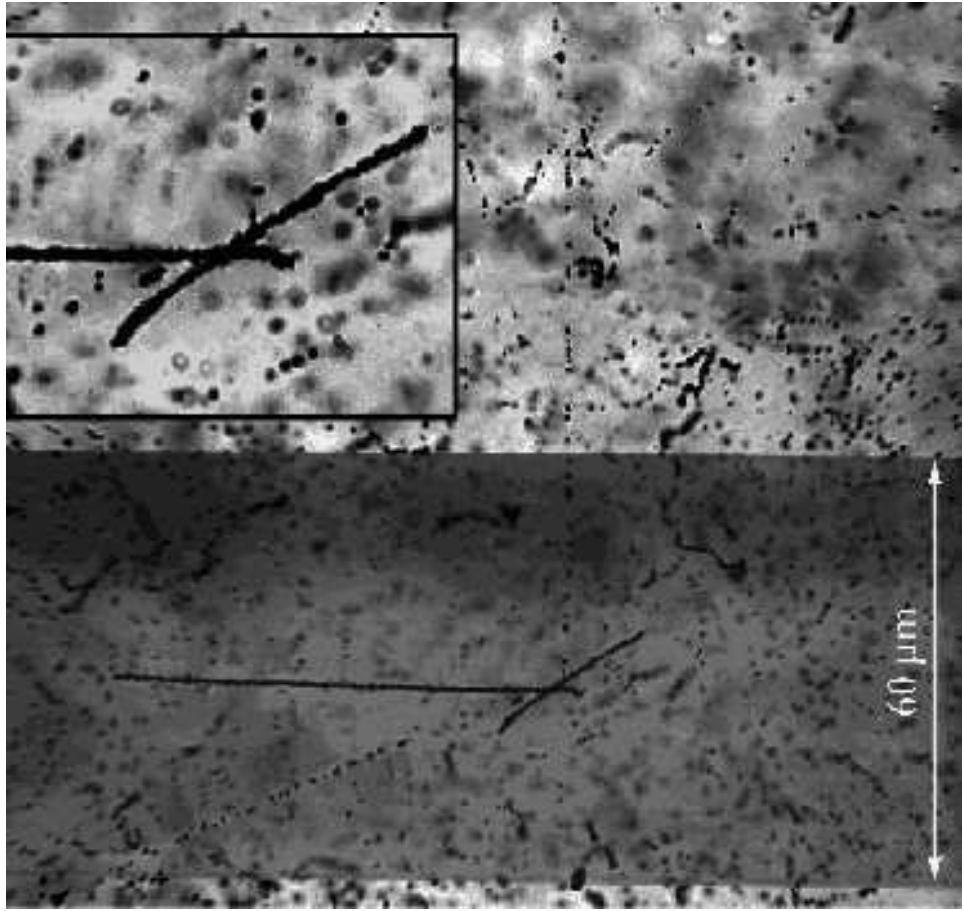}
\caption{ \label{fig:2} Mosaic macrophotography of a hammer-like decay of $^8$He nucleus (horizontal track) stopped in nuclear track emulsion.
 Pair of electrons (point-like tracks) and pair of $\alpha$-particles (short opposite tracks). On insertion (top): enlarged decay
 vertex. To illustrate spatial resolution the image of the decay is superimposed to macrophotography of a human hair
 of thickness of 60 $\mu$m.}
 
\end{figure}

 \indent When scanning the NTE pellicle with a 20$\times$ objectives on the microscopes MBI-9 a primary search for $\beta$-decays of $^8$He nuclei was focused
 on hammer-like events (Fig.~\ref{fig:2}). The absence of tracks of the decay electrons in the event was interpreted as a consequence of an incomplete
 efficiency of observation. Often, in the events named \lq\lq broken\rq\rq ones  gaps were observed between stopping points of  primary tracks and  subsequent hammer-like decays.  In total 1413 "whole" and 1123 "broken" events were found. Decay vertices of 580 "broken" events were found to be laying in a backward hemisphere with the respect  to arrival directons of ions. 
 Corresponding to a half of the statistics this number indicate that the forward-backward asymmetry is absent. The "broken" events were attributed to a drift of thermalized $^8$He atoms that arose as a result of neutralization of $^8$He nuclei. 
This effect is determined by the nature of $^8$He and such events are identified them particularly reliably. \par
 \indent The coordinates of stopping pointsof the ions $^8$He (as well as arrival directions), the decay vertices and stops of decay particles were determined for \lq\lq hammers\rq\rq of 136 \lq\lq whole\rq\rq and
 142 \lq\lq broken\rq\rq events. In \lq\lq broken\rq\rq events the decay points were determined by extrapolating the electron tracks. The
 emission angles and the ranges of $\alpha$-particles were obtained on this basis. The distribution of the opening angles of $\alpha$-particle pairs has a
 mean value $<\Theta_{2\alpha}>$ = (164.9 $\pm$ 0.7)$^{\circ}$ at rms = (11.6$ \pm$ 0.5)$^{\circ}$. Some kink of \lq\lq hammers\rq\rq is defined by the
 momenta carried away by e$\nu$-pairs. The dependence of the $\alpha$-particle ranges L$_{\alpha}$ and their energy values are determined by spline
 interpolation of calculations in the SRIM model \cite{Ziegler}. The mean value of the $\alpha$-particle ranges is (7.4 $\pm$ 0.2) $\mu$m at rms (3.8 $\pm$ 0.2)
 $\mu$m corresponding to a mean energy $<$E($^4$He)$>$ = (1.70 $\pm$ 0.03) MeV at rms 0.8 MeV. Correlation of ranges L$_{1}$ and L$_{2}$ of $\alpha$-particles
 in pairs is clearly manifested. The distribution of the range differences L$_{1}$ - L$_{2}$ has rms 2.0 $\mu$m. \par
\begin{figure}
\centering
  \includegraphics[width=5in]{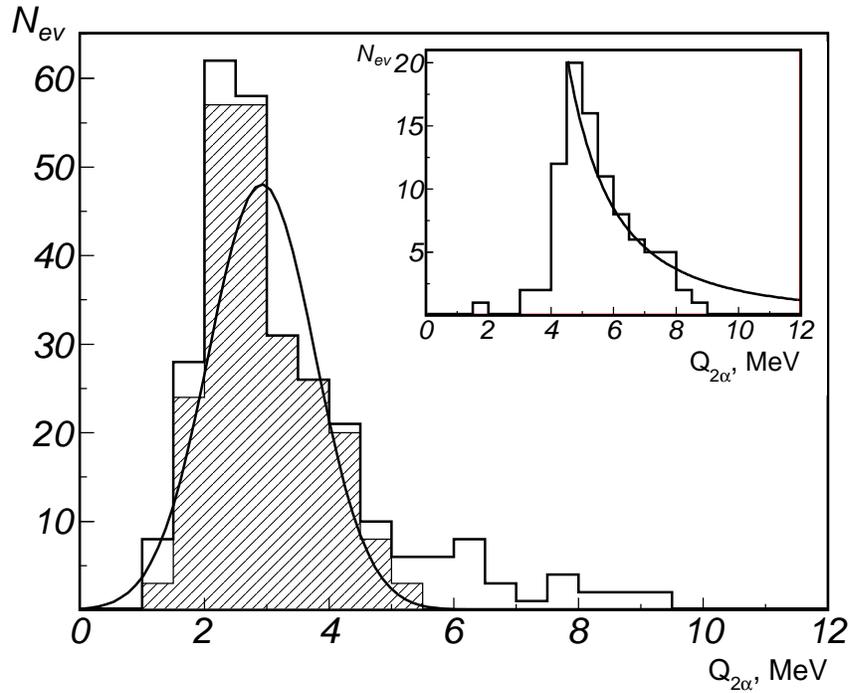}
\caption{ \label{fig:3} Distribution on energy Q$_{2\alpha}$ of 278 pairs of $\alpha$-particles; hatched histogram correspond to condition of
 selection of events L$_{1}$ and L$_{2} < $12.5 $\mu$m, $\Theta > $ 145$^{\circ}$; line - Gaussian. On the insertion: Q$_{2\alpha}$ distribution
 of additional 98 $\alpha$-pairs having L$_{1}$ and L$_{2}  >$ 12.5 $\mu$m.}
    
\end{figure}

\indent Knowledge of the energy and emission angles of $\alpha$-particles allows one to derive the energy distribution of $\alpha$-decays Q$_{2\alpha}$.
 The relativistic-invariant variable Q is defined as the difference between the invariant mass of a final system M$^\star$ and the mass of a primary nucleus
 M, that is, Q = M$^\star$ --M, M$^\star$ is defined as the sum of all products of the 4-momenta P$_{i, k}$ of fragments, that is, 
 M$^{*2}$ = $\sum$(P$_{i}$$\cdot$P$_{k}$). In general, the distribution of Q$_{2\alpha}$ (Fig.~\ref{fig:3}) corresponds to the $^8$Be decay
 from the first excited state 2$^+$. However, the mean value $<$Q$_{2\alpha}>$ is slightly higher than expected. This fact is determined by the presence
 of a "tail" of large values Q$_{2\alpha}$, obviously not matched the description by a Gaussian function. Application of the selection criteria for
 ranges L$_{1}$ and L$_{2}$ less than 12.5 $\mu$m and opening angles $\Theta >$ 145$^{\circ}$, provides a value $<$Q$_{2\alpha}>$ = (2.9 $\pm$ 0.1) MeV
 at RMS (0.85 $\pm$ 0.07) MeV, which corresponds to 2$^+$ state. Ranges  L$_{1}$ and L$_{2}$ stay to be well correlated above 12. 5 $\mu$m. Therefore,
 enhanced ranges  L$_{1}$ and L$_{2}$ can not be attributed to fluctuations of ranges or recombination of ions He$^{+2}$. \par 
 \indent The targeted measurements are continued
 to saturate statistics in the high energy \lq\lq tail \rq\rq Q$_{2\alpha}$ and  to establish its shape. The insertion in Fig.~\ref{fig:3} shows Q$_{2\alpha}$ with additional 98 $\alpha$-pairs having  L$_{1}$ and L$_{2}$ above 12.5 $\mu$m. It should be noted
 that the highly energetic $\alpha$-pairs are among better measurable ones despite to relatively rare appearance.   The physical reason for the appearance of the  "tail"
 in the distribution  Q$_{2\alpha}$ is not clear. Probably, its shape will allow one to verify calculations of spatial structure of 8-nucleon ensembles
 emerging as $\alpha$-pairs of  decays via the state $^8$Be$_{2+}$ \cite{Wiringa}.\par
\begin{figure}
\centering
  \includegraphics[width=5in]{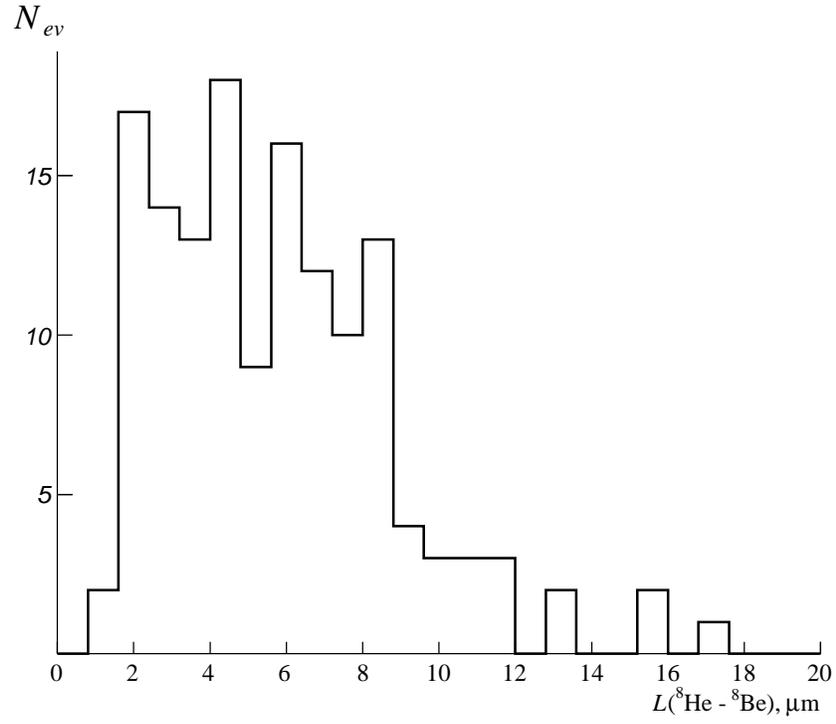}
\caption{ \label{fig:4} Distribution of the distances L($^8$He-$^8$Be) between the stopping points of the $^8$He ions and the decay vertices in the
broken events.}
     
\end{figure}

 \indent In the 142 \lq\lq broken\rq\rq events the distances L($^8$He-$^8$Be) between the stopping points of the $^8$He ions and the decay vertices as well as the angles $\Theta$($^8$He-$^8$Be) between directions of arrivals of the ions and directions from the stopping points of the ions towards the decay vertices 
are defined (Fig.~\ref{fig:4}). Uniformity of distributions of events over these parameters and absence of a clear correlation indicate on on a thermal  drift of the atoms $^8$He. The mean value $<$L($^8$He-$^8$Be)$>$ amounting to (5.8 $\pm$ 0.3) $\mu$m at 
  (3.1 $\pm$ 0.2) $\mu$m, can be associated with a mean range of atoms $^8$He. The low value of a mean speed of the atoms $^8$He defined as ratio of $<$L($^8$He-$^8$Be)$>$ to the half-life of the nucleus $^8$He supports a pattern of diffusion. \par
\indent Observation of the diffusion points to the
 possibility of generating of radioactive atoms $^8$He and pumping them out of sufficiently thin targets. Increasing of the mean speed and
 drift length is achievable due to heating of the target. There is a prospect of accumulating of a significant amount of $^8$He atoms.
 In particular, a $^8$He radioactive gas can be used to measure the half-life of the $^8$He nucleus at a new level of precision and for
 laser spectroscopy of this isotope. Applied interest consists in studies of thin films by pumping atoms $^8$He and their deposition on
 $\alpha$-detectors. Such opportunities are developing intensively with respect of the $^6$He isotope \cite{Knecht}, \cite{Stora}.\par
 \indent To conclude, the result of this work is the demonstration of the opportunities of recently reproduced nuclear track emulsion in a way
 of exposure in a beam of $^8$He nuclei. Nuclear track emulsion made it possible to identify decays of stopped $^8$He nuclei, estimate
 possibilities of $\alpha$-range spectrometry and observe the drift of $^8$He atoms. The high quality of the beam of radioactive
 $^8$He nuclei at the ACCULINNA fragment separator was confirmed. The presented analysis of the decay of
 nuclei $^8$He can serve as a prototype for studying the decays of stopped nuclei $^{8,9}$Li, $^{8,12}$B, $^9$C and $^{12}$N. Statistics of
 hammer-like decays found in this study is a small part of the flux of $^8$He nuclei, and the measured decays - about 10\% of this share.
 This limitation is defined by \lq\lq reasonable expenses\rq\rq of human time and labor. However, nuclear track emulsion in which radioactive nuclei
 are implanted provides a basis for the application of automated microscopy and image recognition software, allowing one to rely on
 unprecedented statistics. Thus, a synergy of classical technique and modern technology can be achieved. This work was supported by the
 grants 12-02-00067, 11-02-00657 and 11-02-00657a of the Russian Foundation for Basic Research and grants of Plenipotentiary representatives of Bulgaria,
 Egypt and Romania at JINR.\par
 

\end{document}